\documentclass[aps,twocolumn,twoside,floatfix,nofootinbib,a4paper]{revtex4}
\usepackage{graphicx}
\usepackage{marginnote}
\usepackage[extra]{tipa}
\usepackage{hyperref}
\usepackage{mathrsfs}
\usepackage{enumitem,kantlipsum}
\usepackage[usenames,dvipsnames]{color}
\usepackage{amsmath}
\usepackage{amssymb}
\usepackage{amsfonts}
\usepackage{mathrsfs}
\usepackage{bm}
\usepackage[all]{xy}
\usepackage{hyperref}
\usepackage{tikz-cd}

\newcommand{\beq}{\begin{equation}}
\newcommand{\eeq}{\end{equation}}
\newcommand{\bea}{\begin{eqnarray}}
\newcommand{\eea}{\end{eqnarray}}
	
\newcommand{\ea}{\end{align*}}

\newcommand{\bma}{\begin{pmatrix}}
\newcommand{\ema}{\end{pmatrix}}

\usepackage[percent]{overpic}
\usepackage{overpic}

\begin{document}
\title{Black holes as portals to an Euclidean realm}
\author{Fan Zhang} 
\affiliation{Institute for Frontiers in Astronomy and Astrophysics, Beijing Normal University, Beijing 102206, China}
\affiliation{Gravitational Wave and Cosmology Laboratory, School of Physics and Astronomy, Beijing Normal University, Beijing 100875, China}

\date{\today}

\begin{abstract}
\begin{center}
\begin{minipage}[c]{0.9\textwidth}
Motivated by a one-cycle cosmological scenario where the big bang marks the egress from a Euclidean metric signature regime, we investigate the possibility of black holes (BH) hosting the mirroring entryways. Analogous to the inflationary stage following the exit end, the entry portals must be enveloped by de Sitter cores inside BHs, in order to satisfy regularity conditions at the metric signature transition boundary. 
We examine the interior structure of BHs that could be consistent with such a physical picture, and conclude that the presence of a spacelike shell of non-inflationary matter is likely required. 
 \end{minipage} 
 \end{center}
  \end{abstract}
\maketitle

\raggedbottom
\section{Introduction and motivation} \label{sec:Motivation}
Singularities at the center of black holes (BHs) signal a breakdown of General Relativity, a problem that had commonly been assumed to be remediable by bringing in quantum mechanics. How this can be achieved is not clear though, and there had even been arguments, termed the principle of unanimity \cite{1977GReGr...8..713W}, asserting that if classical solutions generically exhibit singular behavior, then quantization shouldn't be expected to remove it. One should then at least contemplate the alternative strategy of trying to understand what singularity implies geometrically at the classical level\footnote{This is not to say that quantum considerations do not provide useful hints. For example, if one subscribes to the notion that particles correspond to localized geometric or (differential) topological features of the spacetime fabric itself (e.g., \cite{PhysRev.97.511,2022FoPh...52...46Z}), then it is not difficult to imagine that the usual classical description provides a fully modularized depiction that accounts for only integrated wholistic features like topological charges, but ignores the detailed structures of the particles' innards. Quantum mechanics then tries to offer some additional acknowledgement of internal structures, which is necessary when we zoom in on the particles with magnifying glasses. Going one step further then, with ultracompact matter such as those found at the center of black holes, it is possible that the innards of particles are forced to overlap and combine coherently into large length scales, so a classical (macroscopic) but unconventional equation of state can be evoked to describe it phenomenologically. One likely feature of such an equation of state would be the violation of positive energy conditions that enforce attractive gravity. This possibility is hinted at by quantum fields, that already exhibit tendencies to violate some of the stronger energy conditions, signaling that these conditions may be integral in nature, thus may not be respected when modularization is only partial (quantum) or even not meaningful at all (ultracompact).}, instead of simply sweeping the problem under the rug and wait for quantum gravity to solve it. 

Recently, it had been argued \cite{galaxies8040073} that General Relativity may be based on equiaffine geometry, which breaks down at degenerate points of the extrinsic curvature of our universe, seen as a membrane resting inside of an ambient bulk. If this is true, then there is nothing dramatic in terms of geometry at such degenerate points, and the divergences are only due to the esoteric limitations of the mathematical tools employed (namely, the second fundamental form becomes non-invertible when one or more principal curvatures vanish), and so the generic tendency for General Relativity to develop singularities is not at all a fundamental problem, even at the classical level. It was further shown in \cite{galaxies8040073} that such degenerate points typically arise at the boundary between spacetime regions differing in metric signature (principal extrinsic curvature values, if vary continuously, would naturally hit zero on their way to the opposing sign), with the cosmic big bang being an example. It is then natural to ask whether BH singularities arise for a similar reason, in which case they do not need to be removed (i.e., we do not seek regular BH solutions), just appropriately characterized. 

In this work, the BHs' curvature singularities are proposed to be replaced by Cauchy horizons that mark the boundary of a Euclidean regime, coporealizing, albeit in a distinct fashion, some earlier speculations about metric signature changes in high curvature conditions such as \cite{1984JETP...60..214S}, which are further supported by the fact that there is no built-in mechanisms during gravitational collapse that prevents signature changes \cite{1997JMP....38.4202D}.
One may also understand the generality of signature changes by noting that it is in fact a rather \emph{ad hoc} imposition to force a radial and the temporal eigenvalues of the metric ($A$ and $B$ in Eq.~\ref{eq:zeroshift} below) to strike zero and then switch sign simultaneously in order to maintain signature\footnote{It has been noted by \cite{2003PhRvD..68d4003H} that this is too constraining, forcing them to allow metric signature changes when constructing regular BH solutions.}. 

In fact, if we further specialize our cosmic view, the BHs, thus modified, may in fact serve an essential role in the mechanics of a one-cycle cyclic universe \cite{2022FoPh...52...46Z, Zhang:2019rrc}: the Cauchy horizons are three dimensional hypersurfaces that if we cross, take us into a Euclidean part of our universe, possibly straddling between the future end of our normal Lorentzian universe and the past big bang portal. In this sense, the BHs are quite literally short cuts to the end of the (Lorentzian) world. Such a Euclidean stratum to our overall universe, that BHs take us into, is intuitively appealing since a closed \footnote{Note, despite the overall universe being closed in the ``time'' direction, the ill effects of closed timelike curves are cured by the fact that the propagation of fields in the Euclidean sector is not causal.} large scale topology like a $S^1\times S^3$ brane would be a bit easier to embed into a flat ambient if there are regions where the extrinsic curvature is of a definite signature (visualize the outer rim of a donut). 

The echoing, between the BH and the big bang as twin portals to the Euclidean realm, must go further. 
While we recognize that divergence in curvature tensors could be due to the inapplicability of equiaffine geometry at degenerate points, rather than actual geometric pathology, so we have no physical ground to force perfect regularity on the curvature invariants\footnote{In particular, this means our investigation diverges from \cite{2003PhRvD..68d4003H} that also invoke Euclidean cores in BHs. The underlying constraint for them is that curvature invariants are all regular, leading to e.g., asymptotic flatness (Minkowski or flat Euclidean) at the origin.}, this does not mean we do not impose any regularity condition whatsoever. Just like a coordinate singularity cannot just blow up in just any arbitrary way, if it is to be removable by a change of gauge, the metric signature transition boundary (denoted $\Sigma$ below) should be as regular as possible. Specifically, similar to the big bang side, the core region abutting $\Sigma$ should be de-Sitter\footnote{Recall from \cite{Zhang:2019rrc}, specifically its condition $\mathcal{C}_s^1$ that to ensure a regular extrinsic curvature for $\Sigma$, we need the derivatives (up to second order) of the spatial metric to decline as quickly as the spatial metric itself, which favors a de Sitter-like solution giving an inflation of the spatial scales when moving away from $\Sigma$.} \cite{Zhang:2019rrc} (cf.,  \cite{2024PhRvD.109j4060C} for an example configuration that does not require this condition; in general, the singularity-avoidance property of metric signature changes is well-appreciated in literature, see e.g., \cite{1983PhRvD..28.2960H}), and so there are deflationary segments just prior to the BH entryways to the Euclidean realm, mirroring the big bang side's inflationary stage. In other words, our enlistment of metric signature changes forces a character change to the approach towards singularities, turning from the Kasner-type unbridled chaotic and anisotropic collapse in vacuum, into an isotropic\footnote{For the big bang side, this smoothing necessarily wipes out Weyl curvature \cite{Zhang:2019rrc}, giving rise to a low initial generalized entropy for our early universe \cite{2010JPhCS.229a2013T}. For the BH side though, there is a crucial difference, since a single BH is not a closed thermodynamic system like the earlier universe in its totality, so the overall entropy doesn't need to be destroyed by the smoothing, and can instead reside mostly on the exterior of the BHs. If such entropy, corporealized in Weyl curvature, is indeed the true identity of dark matter \cite{galaxies7010027},  then the proliferation of BHs (possessing more entropy than other celestial objects with similar masses) in late stage universe will provide the necessary attraction to close the universe back up into a big crunch (i.e., the apparent ``weakening'' of dark energy \cite{2024arXiv240403002D} would be explained by the strengthening of its competing dark matter instead),  as preferred by the aforementioned one-cycle scenario (on the other hand, an abundance of primordial black holes during early cosmic history would run counter to the ``increase of entropy'' theme, thus is disfavoured). } de Sitter deflation moderated by inflationary\footnote{Due to e.g., the self-regulatory effect of vacuum polarization \cite{1988CQGra...5L.201P} or restoration of gauge symmetry to the de Sitter group \cite{1992GReGr..24..235D}, or the existence of some upper limit to curvature \cite{Markov1982,2006PhLB..639..368D}.} ultra-dense matter\footnote{Weyl curvature, able to survive in vacuum, contains the off-diagonal differential-between-directions entries in the Riemann curvature (cf., the sectional curvature perspective \cite{1987GReGr..19..771H}), thus is inevitably associated with wild anisotropic behavior when we are approaching a singularity while traversing vacuum. When matter is present however, the Ricci curvature, which represents the diagonal average-across-directions entries, can take over the task of blowing up en-route to singularity, thus facilitates an isotropization of the divergence.} \footnote{Having the inflaton being merely (the ingredients of) ordinary matter all squashed tightly together has the advantage that we no longer need to contrive complicated mechanisms to terminate inflation and encourage reheating. Instead, inflaton would automatically disintegrate into ordinary matter as soon as inflation had escalated the proper spatial volume of comoving neighborhoods sufficiently. In particular, this reaction is forcibly pushed through by a strongly scale-factor-dependent chemical potential, relieving us of having to evoke violent mechanisms such as parametric resonances to catalyze it, which could produce exotic cosmic defects \cite{1998PhLB..440..262T} or strong primordial gravitational waves that are yet to be observed.} \cite{1966JETP...22..241S,1966JETP...22..378G}, which is identified with he cosmic inflaton, existing also in an ultra-dense state\footnote{E.g., in dense nuclear matter, pressure $P$ and energy density $\rho$ are related by $P=n_b(d\rho/dn_b)-\rho$, where $n_b$ is baryon number density and an inflaton-like equation of state is signified by the derivative term vanishing (assuming this expression carries into regimes far beyond nuclear density), or energy density not being sensitive to particle number density. In the alternative quintessence field style approximation, this means the field energy should be independent of field strength, which is possible if the energy is dominated by a flat potential (determined by the binding energy contribution to $\rho$). In particular, an ultra-dense state corresponds to the ultra-relativistic limit, so the pesky rest mass contribution (proportional to the square of the field strength, not independent of it) becomes negligible. In other words, ultra-density is conducive to an inflaton-like state. What remains is to remove the kinetic energy (variations in field strength) which also tends to scale with particle density. This can be achieved if the particles form a highly degenerate amalgam thus cannot be localized. This could plausibly obtain if the particles get crushed so hard against each other that their innards all mesh together, a scenario also compatible with ultra-density.}. We now investigate two configurations of the BH interior that could realize such a scenario, where a metric signature transition boundary $\Sigma$ is hidden behind a de Sitter deflationary core.

\section{Stitched solution} \label{sec:Stitching}
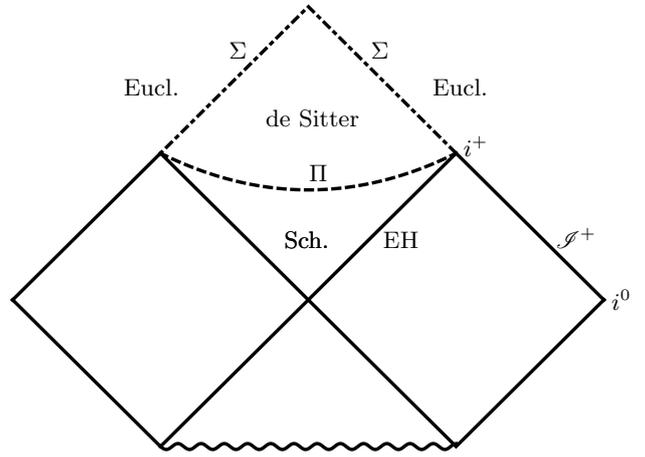
\begin{figure}
\begin{overpic}[width=0.9\columnwidth]{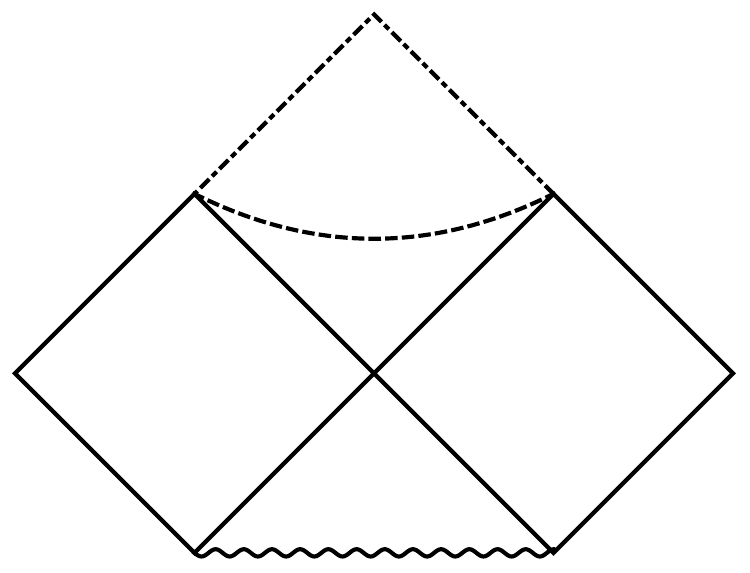}
\put(50,46){{$\Pi$}}
\put(37,66){{$\Sigma$}}
\put(60,66){{$\Sigma$}}
\put(62,35){{EH}}
\put(75,50){$i^+$}
\put(99,25){$i^0$}
\put(90,35){$\mathscr{I}^+$}
\put(43,55){{de Sitter}}
\put(46,35){{Sch.}}
\put(46,35){{Sch.}}
\put(70,60){{Eucl.}}
\put(20,60){{Eucl.}}
\end{overpic}
\caption{The Penrose conformal diagram for the stitched signature change BH. Note the past singularity will be replaced by stellar interior in realistic astrophysical settings. The Euclidean region rests beyond $\Sigma$ but since it doesn't have a causal structure, we don't plot it on the Penrose diagram. Each point on this diagrams is a $S^2$ shell and the dynamical system trajectory of Sec.~\ref{sec:Dyn} below can be seen as a spacelike path turning into a timelike one at the event horizon (hereafter abbreviated to EH), which specifies the metric and scalar field values at the spherical shells that it passes through.}
\label{fig:Penrose}
\end{figure}

To pinpoint the location of $\Sigma$, we note that, analogous to the big bang side \cite{Zhang:2019rrc}, $\Sigma$ should be identified with the $\tau \rightarrow -\infty$ boundary of de Sitter, written in flat slicing coordinates as  
\bea
ds^2 = d\tau^2-e^{2\sqrt{b} \tau} ds^2_{\rm E3}\,,
\eea
($ds^2_{\rm E3} \equiv d\rho^2 + \rho^2 d\Omega^2$ being the flat Euclidean 3-space metric), which is the site of spatial metric vanishing, on its way to switch sign. In the conformal Penrose diagram for the complete de Sitter spacetime, which is a square, this correspond to one of the diagonals. There also exists a mirror-reflected flat slicing coordinate system where the north and south poles of the $S^3$ slices (left and right edges of the conformal diagram) are swapped, for which the other diagonal is the spatial metric vanishing null boundary. We take the bottom two segments of both diagonals (respectively described more conveniently with the reflected and unreflected flat coordinates) as our $\Sigma$, which then bound the lower triangular quadrant of the conformal diagram, see Fig.~\ref{fig:Penrose}\footnote{Note that $\Sigma$ sits on the causal future of null and timelike future infinities $\mathscr{I}^+$ and $i^+$, so our discussion regarding mass inflation in Sec.~\ref{sec:Spinning} applies to it as well. Within our context though, this is not so much of a problem, since $\Sigma$ is supposed to be singular with divergent curvature scalars in the first place (the inverse metric already blows up so any scalars obtained by contraction will generally diverge as well).}. 

This lower quadrant can also be depicted under a time reversed static coordinate system for spherically symmetric spacetime, which has the form 
\bea \label{eq:zeroshift}
ds^2 = -B(r)dt^2 +A(r) dr^2 + r^2 d\Omega^2\,,
\eea
where $r$ is the areal radius (area of constant $r$ and $t$ shells are $4\pi r^2$) and $d\Omega^2$ is the standard $S^2$ line element\footnote{We will be considering spherically symmetric solutions so this is the standard two-sphere metric, but the formalism remains valid for the flat or hyperbolic two-space metrics.},
with $B(r) = 1- b r^2$ (cf., Schwarzschild's $B(r)=1-2M/r$) and $A(r)=1/B(r)$.
The null $\Sigma$ boundary is the future cosmic horizon that is analogous to the null infinity of asymptotically flat spacetimes, which looks as though it could be approachable via timelike curves in a conformal diagram, yet this is only possible if those curves are asymptotically null.  
The static coordinates also align the coordinates with the Killing vectors as best possible, and so a constant $r$ spacelike (we are outside of the cosmic horizon) slice would serve perfectly as the de Sitter versus Schwarzschild boundary $\Pi$ (see Fig.~\ref{fig:Penrose}). This set up is then very similar to the popular universe-inside-a-black-hole scenario proposed by \cite{1990PhRvD..41..383F}, for which $\Pi$ had been shown to be perturbatively stable \cite{1990PhRvD..41..395B}. We differ from this model however, in that we don't extend the lower quadrant triangle into the complete de Sitter square, but sew it onto Euclidean regions instead, so there are no other universes inside of BHs, but instead pathways back to the beginning of our own one-cycle universe.  

Worth noting is the fact that, if one would like to directly stitch an exact de Sitter solution onto an exact Schwarzschild solution as in \cite{1990PhRvD..41..383F,1990PhRvD..41..395B}, the inflaton core (this is customary nomenclature, even though the de Sitter part lives in the temporal future) of BHs are necessarily coated by a spacelike layer of non-inflationary matter \cite{1985NCimL..44..177G} situated at $\Pi$ in order to satisfy the Israel junction conditions \cite{1988CQGra...5L.201P}. 
Such a setup is technically possible to the extent that we allow for some irregularities at $\Pi$, but is not ideal. Firstly, the non-inflationary matter comes into existence suddenly, when its past consists of pure vacuum (the \verb!Sch.! sector in Fig.~\ref{fig:Penrose}), so it cannot quite be interpreted as representing an intermediate stage of the phase transition of ordinary matter into the ultra-dense inflaton. Secondly, the outer solution is exactly Schwarzschild and the soldering surface $\Pi$ is hidden deep inside of the event horizon (EH). There is thus no discernible difference between it and a standard BH for any observer residing outside the EH. In other words, we can not hope for direct observational evidence supporting or refuting our proposal coming from BH measurements. One might try to turn to indirect evidences, including those constraints placed on the inflaton potential by the requirement that such de-Sitter cored BHs can exist, hoping they be verified by cosmological observations. Yet, because there is no dynamics for the inflaton in the stitched solution, we can say very little on this front either.  

\section{Blended solution} \label{sec:Blending}
To cure these shortcomings, we have to excavate more details about the metric, and to do so, we must flesh out the matter source and make it slightly more realistic. We propose that the classical single-field scalar inflaton constitutes the crudest of a phenomenological description for the ultra-dense phase of matter, within which particles lose their individuality and merge into a smeared out amalgam. 
On the other hand, a minimally coupled scalar field also manifests the equation of state of a perfect fluid \cite{1985Ap&SS.113..205M}, offering up a full range of effective equations of state from $p=-\rho$ in the potential-dominated case to $p=\rho$ in the kinetic-dominated regime ($\mathcal{V}$ being the bulk viscosity that determines which), and could thus plausibly also serve as a crude approximate of less extreme phases of matter during different stages of gravitational collapse, filling up in particular the \verb!Sch.! region of Fig.~\ref{fig:Penrose}. In such an optimistic scenario, by examining the Einstein-Klein-Gordon equations, we may be able to better resolve the phase transition and thus smoothen out the isotropization boundary $\Pi$ that had been treated as a shock-like sudden pivot in the stitching solution of Sec.~\ref{sec:Stitching}. 

In other words, if we see the stitching solution as the lowest order approximation, then we could possibly reach the next order up by allowing the scalar field to become dynamic (rather than being fixed giving a constant potential in a perfect de Sitter core, as well as a constant zero potential outside) and omnipresent. Hopefully, deviations from perfect Schwarzschild outside EH could show up at this order, allowing for some observational predictions. 
Such an expectation is inspired by the high profile regular BH solutions in previous literature, such as the Bardeen solution \cite{1968Bardeen}. These are only asymptotically de Sitter and Schwarzschild, therefore do not need the shell and can be everywhere smooth\footnote{On the other hand, gravastars \cite{2001gr.qc.....9035M} also have this type of layering of matter, but there, the mass shell coincides with both the Schwarzschild and de Sitter horizons and is meant to eliminate them so as to evade the information paradox.}. In other words, for them, the near-event-horizon geometry away from either asymptote, could well be only approximately Schwarzschild, casting a slightly modified shadow \cite{2019Univ....5..163D}\footnote{Also, with direct matching solutions that go so far out as to remove the EH, such as gravastars, the ringing spectrum under perturbation could potentially be observable \cite{2009PhRvD..80l4047P}.}.

Unfortunately though, we cannot just grab one of the existing solutions, since they are looking for perfectly regular BHs with a regular origin. For us, the origin is replaced by a $\Sigma$ boundary that doesn't need to be perfectly regular. Therefore, we need to find something similar, yet substantively modified. The right tool for accompanying this task is the dynamical systems approach \cite{2017CQGra..34l5014C,2003CQGra..20.3855C}, which had been shown to be effective and efficient in reproducing, categorizing, and summarizing the regular BH type of solutions. In particular, while works on massless scalar field solutions (i.e., $\mathcal{V}=0$) had been in existence for a long time \cite{Fisher1948,1957PhRv..107.1157B,1968PhRvL..20..878J,1981PhRvD..24..839W}, the more recent arbitrary potential studies had really became possible with the dynamical approach \cite{2017CQGra..34l5014C}. 
This method also has the advantage that it distills the most generous conditions\footnote{For one example of a similar type of condition, recall that the results of \cite{1972PhRvL..28..452B} prevents any scalar hair outside of the horizon (one can still play with the inside or situations where there are no horizons) 
% the proof is regularity of the scalar field between horizon and infinity, does not care about what happens inside horizon. 
when $\varphi \, d\mathcal{V}(\varphi)/d\varphi \geq 0$, or $d^2\mathcal{V}(\varphi)/d\varphi^2 \geq 0$, 
so the metric is necessarily that of Schwarzschild according to Birkhoff's theorem \cite{2005GReGr..37.2253J}. However, if these conditions are violated, we may be able to discern the deviations of the blended solution from Schwarzschild. The stitching solution on the other hand, is already exactly Schwarzschild, so not constrained by this condition.} on $\mathcal{V}$ that could allow for our desired properties, thus informs on how to minimally (i.e., not unnecessarily, thus meaningfully) constrain the inflaton, which had otherwise enjoyed notoriously uninhibited variability that degrades the statistical significance of its ability to match cosmological observations\footnote{An uncharitable analogy would be to try to fit a curve with a variety of different sets of spectral basis functions. They can all do a good job. Although one may argue that parsimonious basis choices that require less free parameters are more natural, it would be even better if we have additional criteria from elsewhere (e.g., the existence of Euclidean portal BHs) that eliminates some candidates from the remaining still vast landscape.}. In contrast, it would be difficult to excise features of the potential associated with any specific solution that are esoteric to the particularities of that solution, rather than being absolutely essential for the qualitative physical features of the Euclidean gateway. 
 
As well, since the scalar matter solution can also be seen as the Einstein frame vacuum solution for modified gravity, and that scalars in the form of moduli generally falls out of higher dimensional tensors whenever extra dimensions are evoked and then compactified like in string theories, the technical aspects of our dynamical systems analysis may be of interest even if the physical picture of an inflaton core does not appeal to the reader.

\subsection{Dynamical systems} \label{sec:Dyn}
The single-scalar inflaton satisfies the Klein-Gordon equation with potential 
\bea \label{eq:KG1}
\nabla^a \nabla_a \varphi - \frac{d \mathcal{V}(\varphi)}{d \varphi} =0\,,
\eea
and has a stress energy tensor of the form 
\bea
T_{ab} = \nabla_a \varphi \nabla_b \varphi - \frac{1}{2}g_{ab}\left( \nabla^c \varphi \nabla_c \varphi + 2 \mathcal{V}(\varphi)\right)\,,
\eea
which satisfies the null energy condition. When in a static spherically symmetric spacetime with metric \eqref{eq:zeroshift}, we end up with Eq.~\eqref{eq:KG1} becoming, 
\bea
\text{KG:} \quad &&2 \varphi'' - \left( A'+B'+r(\cos2\theta-3)\right) \varphi' \notag \\&&- 2 A \frac{d \mathcal{V}(\varphi)}{d \varphi} = 0\,, \label{eq:KG}
\eea
where prime denotes derivative against $r$. The nontrivial entries for the Einstein equations (letting the proportionality constant of the Einstein equation be absorbed by a constant rescaling of $\varphi$ and $\mathcal{V}$), which is required to be valid in either metric signature regimes (the form of the equiaffine Gauss equation is universal whatever the detailed geometry), are 
\begin{align}
\text{Et:} \quad& 2 r A'-2 A^2 \left(r^2 \mathcal{V}(\varphi )+1\right)-A \left(r^2 \varphi'^{2}+2\right) =0\,, \label{eq:Ein1}\\
\text{Er:} \quad& B \left(2 A \left(-r^2 \mathcal{V}(\varphi )+1\right)+r^2 \varphi'^2-2\right)-2 r B' =0 \,, \label{eq:Ein2}\\
\text{E$\theta\phi$:} \quad& A \left(r B'^2-2 B \left(r B''+B'\right)-2 r B^2 \varphi'^2\right) \notag  \\ 
&+ B A' \left(r B'+2 B\right) -4 r A^2 B^2 \mathcal{V}(\varphi )=0 \,. \label{eq:Ein3}
\end{align}
Note that Eq.~\eqref{eq:Ein3} is the common expression shared between the two transverse components ${\theta\theta}$ and ${\phi\phi}$, while the remaining six equations corresponding to the off-diagonal elements of the Einstein and stress-energy tensors are satisfied trivially. 
Eqs.~\eqref{eq:KG} and \eqref{eq:Ein1}-\eqref{eq:Ein3} have the form of a dynamical flow in some parameter space, with $r$ being the only evolution parameter (there are Killing vectors $u^a = B^{-1/2}\delta^a_t$ and tangents to the spherical shells hosting the line elements $d\Omega^2$; this is the reason why we utilize the static coordinates, so that the only explicit dependence is on the non-Killing direction, making the single-variable evolution clear). 

One can further process the evolution system into a first order form more conducive to analysis\footnote{Especially for numerical integrations. In general, it is unreasonable to expect that one's desired solution to a non-linear differential equation like Einstein's to always be analytic, given even those niche tabulated special functions are defined only as eigenvectors to linear differential operators, and are not specifically adapted to the nonlinear problems.}, by defining new variables that are first derivatives of the existing ones along $n^a = A^{-1/2}\delta^a_{r}$. The detailed propagation equations for the static case whereby there exists a timelike Killing vector (this exclude the interior of BHs inside the EH) are given as Eq.~(25) in \cite{2017CQGra..34l5014C}, which we reproduce here
\begin{align}
x_2' &= x_2 x_3 - y_1^2\,,\label{eq:rawevo1}\\
x_3'&=x_2^2+x_2 x_3 - x_3^2 - y_1^2 -1 \,, \label{eq:rawevo2}\\
y_1'&= \frac{\lambda}{\sqrt{2}}\nu - y_1 \left( x_2 + x_3\right)\,, \label{eq:rawevo3}\\ 
\lambda' &= -\sqrt{2}\left(\Gamma -1 \right) \lambda^2 y_1 \label{eq:rawevo4}\,,
\end{align}
where $\nu \equiv x_2^2 +2x_2x_3 - y_1^2 - 1$. 
The physical meanings of the variables are as follows (all the quantities, including the derivatives, are suitably rescaled by the Gauss curvature of the spherical shells at constant $r$ to become dimensionless): $x_2$ is the expansion of the outward radial normal vectors $n^a$ that determines the spherical shells, and carries essentially the information about the local gravitational potential as displayed by $A$ in Eq.~\eqref{eq:zeroshift} \cite{2015CQGra..32j5006G}, so e.g., one typically find $x_2 =0$ at apparent horizons; $x_3$ is the acceleration (along $n^a$) of the $u^a$ that determines the spatial slices, and informs on the radial derivative of the lapse $B$ in Eq.~\eqref{eq:zeroshift}; $y_1$ is the first radial derivative of the scalar field $\varphi$, while the other scalar-field-related quantities are various derivatives of the potential 
\bea
\lambda \equiv -\frac{1}{\mathcal{V}}\frac{d\mathcal{V}}{d\varphi}\,,\quad \Gamma \equiv \left(\frac{d\mathcal{V}}{d\varphi}\right)^{-2} \mathcal{V} \frac{d^2\mathcal{V}}{d\varphi^2}\,,  
\eea 
while $\nu$ (not an independent dynamical variable) is the Gauss curvature rescaled $\mathcal{V}$. 
For a fixed $\mathcal{V}$, $\lambda$ is a substitute for $\varphi$ (cf.~Legendre transformation), and we can then also write $\Gamma(\varphi)$ as a function of $\lambda$ instead. In  \cite{2017CQGra..34l5014C}, a collection of representative potential forms are pre-specified, and the properties of the solutions they permit are then examined. We do the opposite: there is a set of features that we want to generate according to Sec.~\ref{sec:Motivation}, and we want to back out the conditions on $\mathcal{V}$ that enable them. For example, if the streams we end up obtaining have large values of $\lambda$ throughout, while $y_1$ remain constant in sign (so $\varphi$ increases or decreases monotonically), then $\mathcal{V}$ must be steep for significant spans of $\varphi$, making slow-roll inflation less likely. 

We turn now to extend Eqs.~\eqref{eq:rawevo1}-\eqref{eq:rawevo4} into the interior of BHs and also the Euclidean regime. We begin by noting that even though the EH is not a physical singularity, it is nevertheless still a physically significant boundary demarcating two regimes of very different characters, thus could be problematic for the quasi-tetrad modeling approach underlying the dynamical equations, if it depends on said characters. One manifestation of this issue is that inside of the EH, the Killing vector $u^a$ switches from being timelike to being spacelike, so the interior metric cannot be static, but becomes spatially homogeneous. Consequently, instead of the spatial propagation set of Einstein-Klein-Gordon equations, the dynamical system switches over to a temporal evolution. The interior evolution is thus not considered in 
\cite{2017CQGra..34l5014C} (the flows cut off at the EH), and is given as a separate dynamical system in \cite{2015CQGra..32j5006G}. Having in mind an additional adaptation needed for the Euclidean regime though, we adopt an alternative, more unified approach wherein the changes are minimized. Specifically, instead of maintaining the quasi-tetrad formalism in its rudimentary form and tailor it into a different version for each specific situation, we generalize the formalism by complexification, so as to be able to retain the same formal derivation steps, and only specialize by imposing the congruous reality conditions in the end. 

Note first that much of the definitions and projections evoked by the $1+1+2$ (along $u^a$, $n^a$ and the spherical shells) decomposition can be carried over to different metric eigenvalue sign arrangements so long as we change some of the signs in the definitions. For example, the projection operator $h_{ab}=g_{ab} + u_{a}u_{b}$ should really pick up a sign change in the BH interior if we want to keep using $u_a$ as the Killing unit vector. Also, the explicit $1+1+2$ decomposition of the stress-energy tensor for the scalar field as in Eq.~(15) of \cite{2017CQGra..34l5014C}, should in principle have to change as well. The simplest solutions to avoid the hassle however, which we will adopt in this paper, is to simply apply the ``wick rotation'' $u_a = i \nu_a$ (and similarly rotate $n^a$ into being purely imaginary where appropriate), where now $\nu_a$ is the physical unit vector while $u_a$ becomes just some complexified auxiliary vector field without direct physical meaning, but still retain the expressions like $u^a = B^{-1/2} \delta^a_t$ where now we can have $B<0$ and $u^a$ becomes purely imaginary. This ensures all the expressions and equations retain their apparent form, but can become imaginary (we summarize the impact of complexification for the dynamical variables in Tb.~\ref{tb:Complex}), and the price we pay is that the physical interpretations of the $1+1+2$ decomposed quantities, such as pressure or acceleration, which are still based off of $u_a$ rather than $\nu_a$ (all expressions are kept unchanged, including the definitions), requires some refreshing (e.g., the acceleration now picks up a more generic interpretation of being the rate of change of an arbitrary Killing vector rather than the timelike normal). 
Further down stream, The metric-signature-switching boundary $\Sigma$ is yet another singular point for the dynamical system, and we can treat the dynamical system beyond it in a similar way, given that it is also a sign-change related issue. However, since the coefficient to $d\Omega^2$ would also need to change sign, it is more convenient to carry out an extra overall sign reversal for the entire metric in the Euclidean regime, and deal with $(+,+,+,+)$ rather than $(-,-,-,-)$, so that we have $B<0$ but $A>0$ in that regime, equivalently, we only make $u^a$ imaginary. 
 
\begin{table}[t]
\caption{The impacts on various expressions in the dynamical system when $u^a$ ($u^t=1/\sqrt{B}$ \cite{2015CQGra..32j5006G}) and/or $n^a$ ($n^r=1/\sqrt{A}$) picks up an extra $i$ factor at the various boundaries.\\} % title of Table
\centering % used for centering table
\begin{tabular}{c c c} % centered columns (4 columns)
\hline\hline %inserts double horizontal lines
Quantity & $\quad u^a: \Re \rightarrow \Im \quad$ & $\quad n^a: \Re \rightarrow \Im \quad$  \\ 
[0.5ex] % inserts table
%heading
\hline % inserts single horizontal line
$\cdot'$ & $\Re \rightarrow \Re$ & $\Re \rightarrow \Im$ \\ % inserting body of the table
 $x_2$ & $\Re \rightarrow \Re$ & $\Re \rightarrow \Im$ \\
 $x_3$ & $\Re \rightarrow \Re$ & $\Re \rightarrow \Im$ \\
 $y_1$ & $\Re \rightarrow \Re$ & $\Re \rightarrow \Im$ \\
 $\lambda$ & $\Re \rightarrow \Re$ & $\Re \rightarrow \Re$ \\
 $\Gamma$ & $\Re \rightarrow \Re$ & $\Re \rightarrow \Re$ \\ [1ex]
 \hline %inserts single line
\end{tabular}
\label{tb:Complex} 
\end{table}

Furthermore, while the Minkowski asymptotic is a fixed point (${x'_2}={x'_3}={Y'_1}={\lambda'}=0$) for the flow\footnote{\label{fn:Min}That is, trajectories need to get stuck there and terminate. If the flow continues past, then we will have additional spacetime regions outside of the Minkowski shell, so it is not asymptotically Minkowski but only accidentally so at some intermediate radius. Furthermore, globally Minkowski solution, represented by a trajectory stuck at a single point (recall all quantities are rescaled with the Gaussian curvature of the spherical shells, so the dynamical quantities on different shells in a flat Minkowski space would be the same), is also a viable solution, thus the Minkowski point must be a fixed one.} on the regular interior of the phase space, as previously mentioned, many features most interesting to us often coincide with some dynamical variable blowing up, thus exist on the infinite boundary of the phase space. To facilitate further analysis then, it is useful to compactify the phase space and pull the infinity in by evoking the Poincar\'e variables\footnote{A word of caution is that the topology of the phase space, in particular its asymptotic region, can be nontrivial. For example, some combination of the variables blowing up to $+\infty$ may be physically equivalent to when they blow down to $-\infty$, leading to some points on the asymptotic cylinder being identified with each other (see e.g., Fig.~3 in \cite{2015CQGra..32j5006G}). Moreover, we are dealing with a four dimensional phase space, so generically, the differential or smoothness structure is not unique even if we manage to pin down the topology.} \cite{2017CQGra..34l5014C}
\bea
&&X_2 = \frac{x_2}{\xi}\,, \quad X_3 = \frac{x_3}{\xi}\,, \quad Y_1 = \frac{y_1}{\xi}\,, \notag \\ 
&&\xi = \sqrt{1+ x_2^2+x_3^2+y_1^2}= \frac{1}{\sqrt{1-X_2^2-X_3^2-Y_1^2}}\,,  \label{eq:xi}\quad 
\eea
to place the phase space boundary (for $x_2$, $x_3$, $y_1$ and $\lambda$) at the cylinder $\{X_2^2+X_3^2+Y_1^2 = 1\,, \lambda \in \mathbb{R}\}$. 
It is worth noting that the $\lambda$ direction is not compactified, even though $\lambda$ can blow up as well when $\mathcal{V}=0$, e.g., on its way to change sign. However, according to our physical motivation in Sec.~\ref{sec:Motivation}, there must be a mechanism to prevent the spacetime from crumpling up, which would squash the non-trivial geometrodynamic features that constitute particles ever closer even in the ultra-dense configurations, making impossible the equiaffine minimal surface characterization. Only a negative\footnote{Note the sign convention is such that the stress-energy tensor contains the $-g_{ab}\mathcal{V}$ term, so this correspond to an expansive inflaton, however, since this phase of matter only exists in extremely compact objects, this potential cannot explain the dark energy.} $\mathcal{V}$ (and thus a negative $\nu$) can achieve this, so we make this assumption of negativity. The allowed compactified phase space is thus 
\begin{align} \label{eq:PhaseEx}
\{(X_2, X_3, Y_1, \lambda): \,&\mu \equiv 2X_2^2+2X_2X_3+X_3^2 -1 < 0\,,\,\, \notag \\
&X_2^2+X_3^2+Y_1^2 \leq 1\,, \,\, \lambda \in \mathbb{R}\}\,,
\end{align}
with the flow given by (Eq.~58 of  \cite{2017CQGra..34l5014C})
\begin{align} 
\tilde{X_2} =& -\frac{\lambda}{\sqrt{2}}X_2 Y_1 \mu + Y_1^2\left( 2X_2^2 + X_2X_3-1\right)\notag \\
&- X_2X_3\left(3X_2^2+X_2X_3-2 \right)\,,  \label{eq:PropX2} \\
\tilde{X_3} =& -\frac{\lambda}{\sqrt{2}}X_3 Y_1 \mu - 3X_2^2X_3^2+2X_2^2-X_2X_3^3 \notag \\
&+ X_3Y_1^2\left(2X_2 +X_3\right) + X_2 X_3 +X_3^2-1\,, \label{eq:PropX3}\\
\tilde{Y_1} =& \frac{\lambda}{\sqrt{2}}\left(1-Y_1^2 \right) \mu - X_2 Y_1 \left( 3X_2X_3 +X_3^2+1\right) \notag \label{eq:PropY1} \\
&+ Y_1^3\left( 2X_2+X_3\right)\,,\\
\tilde{\lambda} =& -\sqrt{2}Y_1 \left(\Gamma -1 \right)\lambda^2\,, \label{eq:PropLambda}
\end{align}
where $\tilde{\cdot}\equiv \sqrt{1-X_2^2-X_3^2-Y_1^2}\cdot '$. 

We should be cognizant of the fact that when $x_2$, $x_3$ and $y_1$ become imaginary (i.e., in the intervening region between the EH and $\Sigma$), we would need to do something different\footnote{In principle, we could stay with the $X_2$ etc variable system throughout, in which case the trajectory can pierce through the sphere of $X_2^2+X_3^2+Y_1^2=1$ and continue on into its exterior. However, we won't be able to keep evolving smoothly outside that sphere since there is a jump into negative $X_2^2+X_3^2+Y_1^2$ whenever $\xi =0$.}, otherwise $X_2$, $X_3$, $Y_1$ can still diverge (to both $\pm \infty$) at $x_2^2+x_3^2+y_1^2=-1$. Therefore, we define the real variables (marked with a superscript $I$ for ``intervening region'')
\bea 
&&X^I_2 = \frac{-i x_2}{\xi^I}\,, \quad X^I_3 = \frac{-i x_3}{\xi^I}\,, \quad Y^I_1 = \frac{-i y_1}{\xi^I}\,, \notag \\ 
&&\xi^I = \sqrt{1- x_2^2 - x_3^2 - y_1^2}\,,  \label{eq:xiI}
\eea
so once again 
\bea
X^{I2}_2+X^{I2}_3+Y^{I2}_2 = \frac{|x_2^2+x_3^2+y_1^2|}{1+|x_2^2+x_3^2+y_1^2|} \leq 1\,,
\eea
and since we have
\bea
\xi^I = \frac{1}{\sqrt{1-X_2^{I2}-X_3^{I2}-Y_1^{I2}}}\,,
\eea
with $x_2 = i X_2^I \xi^I$ etc, closely mimicking the un-superscripted expressions, the translation of Eqs.~\eqref{eq:rawevo1}-\eqref{eq:rawevo4} into the evolution equations for the $(X^I_2, X^I_3,Y^I_1,\lambda)$ system leads to very similar expressions as Eqs.~\eqref{eq:PropX2}-\eqref{eq:PropLambda}, only with everything on the right hand side (including inside of $\mu$) being altered with the replacement $X_2 \rightarrow i X_2^I$, $X_3 \rightarrow i X_3^I$, and $Y_1 \rightarrow i Y_1^I$. For the left hand side, it is more convenient to redefine $\tilde\tilde{\cdot}\equiv -i \sqrt{1-X_2^{I2}-X_3^{I2}-Y_1^{I2}} \cdot '$ so there is an extra $-i$ factor that annihilates with the $i$ factor accompanying the linear $X^I_2$, $X^I_3$ and $Y^I_1$ appearing in the left hand sides. For $\lambda$ on the other hand, this $-i$ from the derivatives cancel with the $i$ from $Y_i^I$ on the right hand side instead, leaving behind a minus sign. Overall thus, we end up with
\begin{align} 
\tilde{\tilde{X_2}}^{I} =& \frac{\lambda}{\sqrt{2}}X_2^{I} Y_1^{I} \mu^{I} - Y_1^{I2}\left( -2X_2^{I2} - X_2^{I}X_3^{I}-1\right)\notag \\
&+ X_2^{I}X_3^{I}\left(-3X_2^{I2}-X_2^{I}X_3^{I}-2 \right)\,,  \label{eq:IPropX2} \\
\tilde{\tilde{X_3}}^{I} =& \frac{\lambda}{\sqrt{2}}X_3^{I} Y_1^{I} \mu^{I} - 3X_2^{I2}X_3^{I2}-2X_2^{I2}-X_2^{I}X_3^{I3} \notag \\
&+ X_3^{I}Y_1^{I2}\left(2X_2^{I} +X_3^{I}\right) - X_2^{I} X_3^{I} -X_3^{I2}-1\,, \label{eq:IPropX3}\\
\tilde{\tilde{Y_1}}^{I}=& \frac{\lambda}{\sqrt{2}}\left(1+Y_1^{I2} \right) \mu^{I} + X_2^{I} Y_1^{I} \left(- 3X_2^{I}X_3^{I} -X_3^{I2}+1\right) \notag \label{eq:IPropY1} \\
&+ Y_1^{I3}\left( 2X_2^{I}+X_3^{I}\right)\,,\\
\tilde{\tilde{\lambda}} =& \sqrt{2}Y^{I}_1 \left(\Gamma -1 \right)\lambda^2\,, \label{eq:IPropLambda}
\end{align}
where now the allowed phase space region is 
\begin{align} \label{eq:PhaseIn}
\{(X_2^{I}, X_3^{I}, Y_1^{I}, \lambda):\, &\mu^I \equiv -2X_2^{I2}-2X_2^{I}X_3^{I}-X_3^{I2} -1 < 0\,,\,\, \notag \\
&X_2^{I2}+X_3^{I2}+Y_1^{I2} \leq 1\,, \,\, \lambda \in \mathbb{R}\}\,.
\end{align}
This is a different dynamical system due to the disparate sign changes among terms composing of varying powers of the dynamical variables, thus e.g., there isn't a Minkowski fixed point. 

\subsection{Boundaries and horizons} 
We can find the location for $\Sigma$, which is the de Sitter horizon in the static coordinates (see Fig.~\ref{fig:Penrose}), by first recalling the relationship \cite{2015CQGra..32j5006G} 
\begin{align} \label{eq:ToMetric}
A(r) = \frac{1}{x_2^2} \,, \quad  \frac{d \ln B(r)}{d \ln r} = 2 \frac{x_3}{x_2} \,,
\end{align}
which gives, for the de Sitter solution,   
\bea
x_2 = \sqrt{1-b r^2}\,, \quad x_3 = - \frac{br^2}{\sqrt{1-br^2}}\,,  
\eea
so the horizon $r = 1/\sqrt{b}$ corresponds to $x_2=0$ and
\begin{equation}
\left\{ \begin{aligned} 
  &x_3 \rightarrow + i \infty \quad \text{from large $r$ side}\,, \\
  &x_3 \rightarrow - \infty \,\,\quad \text{from small $r$ side}\,,
\end{aligned} \right.
\end{equation}
leading to $\xi = x_3 \rightarrow -\infty$ in Eq.~\eqref{eq:xi} and $\xi^I = -i x_3 \rightarrow  +\infty$ in Eq.~\eqref{eq:xiI}. 
Since we should be deep inside of the de Sitter regime now that we are almost at $\Sigma$, we must also have that $\mathcal{V}$ should be nearly constant, accomplished by either $\varphi$ being roughly constant, i.e., $y_1\approx 0$, or $\lambda \approx 0$ so changes in $\varphi$ does not translate into changes in $\mathcal{V}$. With the latter case, $y_1$ should not diverge either, to ensure regularity at $\Sigma$ \cite{Zhang:2019rrc}, leading to $\Sigma$ at 
\bea \label{eq:SigmaLocE}
X_2|_{\Sigma} = 0\,, \quad X_3|_{\Sigma}=-1\,, \quad Y_1|_{\Sigma}=0\,, 
\eea
when approached from the Euclidean side, and equivalently
\bea \label{eq:SigmaLocL}
X^{I}_2|_{\Sigma} = 0\,, \quad X^{I}_3|_{\Sigma}=1 \,, \quad Y^{I}_1|_{\Sigma}=0\,,   
\eea 
when approached from the Lorentzian de Sitter side (recall that the \verb!de Sitter! region in Fig.~\ref{fig:Penrose} is the outside of the de Sitter horizon thus $r$ is timelike). The condition on $y_1$ is lost when going into the Poincar\'e variables, since finite and zero $y_1$ values are the same when rescaled by the diverging $\xi$ or $\xi^I$, but we should remember that if $y_1 \approx 0$, then $\lambda$ can be anything, or else we must have $\lambda \approx 0$. 

On the other hand, for the Schwarzschild solution, we have according to Eq.~\eqref{eq:ToMetric} that 
\bea
x_2 = \sqrt{1-\frac{2M}{r}}\,, \quad x_3 = \frac{M/r}{\sqrt{1-2M/r}}\,.  
\eea
and the EH at $r = 2M$ corresponds to $x_2=0$ and
\begin{equation}
\left\{ \begin{aligned} 
  &x_3 \rightarrow + \infty\,\, \quad \text{from large $r$ side}\,, \\
  &x_3 \rightarrow - i\infty \quad \text{from small $r$ side}\,.
\end{aligned} \right.
\end{equation}
with $\xi = x_3 \rightarrow +\infty$ in Eq.~\eqref{eq:xi} and $\xi^I = - x_3 \rightarrow  +\infty$ in Eq.~\eqref{eq:xiI}. Once again, regardless of whether $y_1$ vanishes at the EH, so long as it does not diverge, we end up with $Y_1=0$. Subsequently, the EH is at  
\bea \label{eq:HorizonLocEx}
X_2 |_{\rm EH}= 0\,, \quad X_3|_{\rm EH}=1\,, \quad Y_1|_{\rm EH}=0\,, 
\eea
when approached from the exterior of the BH, and equivalently
\bea \label{eq:HorizonLocIn}
X^{I}_2|_{\rm EH} = 0\,, \quad X^{I}_3|_{\rm EH}=-1 \,, \quad Y^{I}_1|_{\rm EH}=0\,, 
\eea 
when approached from the interior. The $\lambda$ value is not constrained, since $\varphi$ makes no explicit appearances in the metric that determines the horizon.  

One should note that Eqs.~\eqref{eq:HorizonLocIn} and \eqref{eq:SigmaLocL} are not fixed points\footnote{The fixed points of the Poincar\'e transformed system are not necessarily fixed points of the original dynamical system \eqref{eq:rawevo1}-\eqref{eq:rawevo4} (e.g., the EH is not a fixed point for the lower case letter propagation and evolution systems in either Sec.~IIIA or B of \cite{2015CQGra..32j5006G}), so there is no physical reason to demand that the various horizons be fixed points in the former. In fact, there are no admissible (recall that the variables are real) fixed points for the intervening dynamical system, as plugging the equations into any symbolic computation package will verify. } of the intervening dynamical system \eqref{eq:IPropX2}-\eqref{eq:IPropLambda}, so there can be only one trajectory passing through each (a corollary of the existence and uniqueness theorems, see e.g., pp 150 of \cite{strogatz2007nonlinear}), and in fact, since they differ only in $X_3^I$, the straight line given by setting \eqref{eq:IPropX2}-\eqref{eq:IPropLambda} with $X^I_2=0$ and $Y^I_1=0$, or explicitly
\begin{align} 
\tilde{\tilde{X_2}}^{I} =&\, 0 \,, \quad   
\tilde{\tilde{X_3}}^{I} = -X_3^{I2}-1\,, \notag \\
\tilde{\tilde{Y_1}}^{I}=& -\frac{\lambda}{\sqrt{2}} (X_3^{I2}+1)\,,\quad
\tilde{\tilde{\lambda}} = \, 0\,, 
\end{align}
is exactly a bridge flowing from $\Sigma$ towards the EH in the increasing $r$ direction, provided that we pick $\lambda =0$ on both ends (and it will remain so through out the transit). This is a rather boring solution however, since $Y_1$ vanishes throughout, thus $\varphi$ is just a constant that never evolves\footnote{Note though this is not a trivial solution, since we don't have a cosmological constant, but rather a dynamical quinessence field that satisfies a nontrivial equation of motion. This is the reason why we don't have the Schwarzschild solution within our dynamical system, since an active scalar field introduced into a black hole won't just sit around at a constant value.}. This is fine for the dense, potential $\mathcal{V}$ dominated core, but as aforementioned, we need to precede it with a kinetic term dominated (in the stress-energy tensor) epoch of tenuous matter comprising the collapsing stellar material. The vanishing kinetic term associated with the straight-line solution clearly fails this physical motivation. 
What we want is instead a solution that resembles the Schwarzschild solution near the EH end, but nevertheless reaches $\Sigma$ eventually. Since at the Schwarzschild singularity, we have that $x_2 \rightarrow i \infty$, $x_3 \rightarrow -x_2/2$, and also $\xi^I \rightarrow (\sqrt{5}/2) \infty$, leading to ($Y_1|_{\rm Sing}$ and $\lambda$ can be anything since we only care about the metric mimicking Schwarzschild)
\bea
X_2|_{\rm Sing}= \frac{2}{\sqrt{5}}\,, \quad X_3|_{\rm Sing} = -\frac{1}{\sqrt{5}}\,, 
\eea
we would need a streamline heading off away from the EH towards this general direction initially. However, this is not possible without some tweaking, since the EH is not a fixed point, and thus the initial directions at it is fixed onto that aforementioned straight line. This ties in with the fact that it is desirable for there to be a shell of non-inflationary matter forming a non-smooth boundary between the (exact) Schwarzschild and de Sitter solutions, which could facilitate an abrupt turning of the trajectory, should it be needed, thus provides an added flexibility to more easily connect up the two desired end points.  

The smoother blended solution we are searching for in this section does not contain such a shell, and the flexibility, if possible at all, has to come from elsewhere. Given we are not obliged to have
exact de Sitter and Schwarzschild regions, only approximate ones, we could in principle have $\Sigma$ and the EH being two locations near but not necessarily precisely at \eqref{eq:HorizonLocIn} and \eqref{eq:SigmaLocL}. In other words, instead of forcing a kink in the trajectory so that it can bridge two rigidly fixed end points, one might hope to keep the trajectory smooth, but allow the ends to float a little. However, this strategy does not appear promising at the level of details afforded by our 1-D dynamical system. To see why, we can linearize the propagation equations near the EH, we get ($\epsilon$ is a flag for the order, i.e., the power of perturbative $\delta$ quantities; $\lambda_0$ is the $\lambda$ value at the EH)
\begin{align}
\tilde{\tilde{\delta X^I_2}}|_{\rm EH}\approx&\, 2  \delta X^{I}_2 + \mathcal{O}(\epsilon^2)\,, \notag \\
\tilde{\tilde{\delta X^I_3}}|_{\rm EH}\approx&\, -2 + (2  \delta X^I_2 + 2\delta X^I_3 + \sqrt{2} \delta Y^I_1 \lambda_0) + \mathcal{O}(\epsilon^2)\,, \notag \\
\tilde{\tilde{\delta Y^I_1}}|_{\rm EH}\approx&\, -\sqrt{2} \lambda_0 +\sqrt{2} \left(-\delta \lambda+(\delta X^I_2+\delta X^I_3)\lambda_0\right)  + \mathcal{O}(\epsilon^2)\,, \notag \\
\tilde{\tilde{\delta \lambda}}|_{\rm EH}\approx&\, \mathcal{O}(\epsilon^2)\,. 
\end{align}
Because the flow from the EH towards either the $\Sigma$ or the Schwarzschild singularity direction would be decreasing in $r$, we should really regard the right hand side of all the linearized equations as being padded by an extra minus sign. Consequently, any initial $\delta X_2^I$ would be exponentially decreasing, and thus it would be very difficult to contrive a solution that resembles the Schwarzschild solution near the EH.

\section{Conclusion} \label{sec:Spinning}
In this exploratory note, we have taken a preliminary look at the BHs serving as portals into Euclidean domains of our universe. A dynamical systems analysis indicates that a smooth transition from the Schwarzschild to the de Sitter regimes in the interiors of BHs (the blended solution) would be difficult if not impossible to achieve, while a stitched solution containing a transitory shell of non-inflationary matter, akin to that seen in the universe-in-a-BH proposals, would be much more readily obtainable. One might view the situation as follows: the stitched solution ignores the collapsing stellar matter, thus inevitably ends up having to introduce a very singular shell popping matter into existence all of a sudden; while the blended solution goes to the other extreme and hopes to not have any sharpness at the phase transition at all, even though all phases of matter are rendered only in the crudest manner, enlisting nothing but a single scalar field in a highly symmetric setting\footnote{One obvious inadequacy due to high symmetry is that the existence of a Killing vector $u^a$ (see Sec.~\ref{sec:Dyn}) implies the outside of the EH is static for us, so we don't have a proper full account of the stellar collapse; the outside region is only an approximate to the late stages of collapse after most matter has crossed into the EH. On the other hand, the inside is homogeneous so the stellar matter that entered the EH is not contracting, but instead undergoes an alteration of its equation of state in situ. Thus, while we may capture the gradualness of the phase transition, the fact that it is triggered and controlled by the gradual increase in compactification is not readily apparent within our simplifying high-symmetry context.}. 
The realistic situation is perhaps somewhere in-between, whereby the shell is smoothable, yet only when one manages to more faithfully account for the microphysics of the phase transition into the ultra-dense inflationary matter (its impact on a dynamical systems analysis like ours would be to introduce more independent terms into the stress-energy tensor, thereby increase the number of dynamical variables, or in other words the dimension of the phase space) and relax the constriction from symmetry (this would introduce more underlying propagation or evolution parameters into the dynamical system, thus substantially alter the nature of the analysis). 

For the latter direction of potential future attack, the most obvious next step is to introduce spin, thus reduce the symmetry from spherical to axial. The most striking new feature for the interior of a spinning black hole is arguably that closed timelike curves can exist behind the inner horizons of Kerr-Newman solutions (so long as they contain some segments protruding into the negative Boyer-Lindquist $r$ regions), or in other words, if we keep the BH region inside of the inner horizon Lorentzian, it becomes very problematic for the unitary evolution of fields (the singularities and the infinities of the infinite chain of parallel universes in the maximally extended solution do us no favors in this regard either). The situation is further complicated but not improved by the fact that there is an infinite blue shift and a resulting mass inflation \cite{1989PhRvL..63.1663P,1990PhRvD..41.1796P} at the inner horizon of the Kerr-Newman solutions, which are notoriously sensitive to global features of the spacetime\footnote{For example, one can easily understand the infinite blueshift blowup in an asymptotically flat universe by noting that the future timelike infinity is null connected to the inner horizon and is thus in its causal past, so the horizon can see information coming in from infinite amount of future time, thus signal must be blue shifted by an infinite amount to compactify them. In more details, the conformal transformation that squeezes the entire regions near the future timelike infinity into a single point on the Penrose diagram must squeeze radiation in those regions to become infinitely blue shifted when this radiation travels to the Cauchy horizon.}\footnote{Changing the asymptotic geometry, e.g., by placing the BH in a de Sitter universe, alters the hole's internal structure \cite{1997AnIPS..13...33C} even though the instability doesn't go away \cite{1998PhRvL..80.3432B}.}.
An optimist might expect the post-mass-inflation singular inner horizon would be able to hide the closed timelike curves behind, but they exert only finite tidal force \cite{PhysRevLett.67.789,1998PhRvD..57.7084B} on observers attempting to cross them, thus are in principle penetrable. One could then argue that the Kerr-Newman solutions already flag an inevitable breakdown of causality, and we might as well do a more thorough job by turning the problematic regions Euclidean and remove even local hyperbolicity. In other words, with spinning BHs, it would be rather natural to expect $\Sigma$ to sit at, or outside of the inner horizon\footnote{Note in particular, beyond the original motivation for their introduction, this incidental fact also sets the Euclidean portal BHs apart from the upper-bound-on-curvature studies that place the de Sitter region within some Planck time or length (depending on whether there is an inner horizon) from the curvature singularity.}. 

\acknowledgements
This work is supported by the National Key Research and Development Program of China grant 2023YFC2205801, and the National Natural Science Foundation of China grants 12433001, 12021003. 

\clearpage
\bibliography{BHSignatureChange.bbl}

\begin{thebibliography}{45}
\expandafter\ifx\csname natexlab\endcsname\relax\def\natexlab#1{#1}\fi
\expandafter\ifx\csname bibnamefont\endcsname\relax
  \def\bibnamefont#1{#1}\fi
\expandafter\ifx\csname bibfnamefont\endcsname\relax
  \def\bibfnamefont#1{#1}\fi
\expandafter\ifx\csname citenamefont\endcsname\relax
  \def\citenamefont#1{#1}\fi
\expandafter\ifx\csname url\endcsname\relax
  \def\url#1{\texttt{#1}}\fi
\expandafter\ifx\csname urlprefix\endcsname\relax\def\urlprefix{URL }\fi
\providecommand{\bibinfo}[2]{#2}
\providecommand{\eprint}[2][]{\url{#2}}

\bibitem[{\citenamefont{{Wheeler}}(1977)}]{1977GReGr...8..713W}
\bibinfo{author}{\bibfnamefont{J.~A.} \bibnamefont{{Wheeler}}},
  \bibinfo{journal}{General Relativity and Gravitation}
  \textbf{\bibinfo{volume}{8}}, \bibinfo{pages}{713} (\bibinfo{year}{1977}).

\bibitem[{\citenamefont{Wheeler}(1955)}]{PhysRev.97.511}
\bibinfo{author}{\bibfnamefont{J.~A.} \bibnamefont{Wheeler}},
  \bibinfo{journal}{Phys. Rev.} \textbf{\bibinfo{volume}{97}},
  \bibinfo{pages}{511} (\bibinfo{year}{1955}),
  \urlprefix\url{http://link.aps.org/doi/10.1103/PhysRev.97.511}.

\bibitem[{\citenamefont{{Zhang}}(2022)}]{2022FoPh...52...46Z}
\bibinfo{author}{\bibfnamefont{F.}~\bibnamefont{{Zhang}}},
  \bibinfo{journal}{Foundations of Physics} \textbf{\bibinfo{volume}{52}},
  \bibinfo{eid}{46} (\bibinfo{year}{2022}), \eprint{2108.05724}.

\bibitem[{\citenamefont{Zhang}(2020)}]{galaxies8040073}
\bibinfo{author}{\bibfnamefont{F.}~\bibnamefont{Zhang}},
  \bibinfo{journal}{Galaxies} \textbf{\bibinfo{volume}{8}}, \bibinfo{pages}{73}
  (\bibinfo{year}{2020}), ISSN \bibinfo{issn}{2075-4434},
  \urlprefix\url{https://www.mdpi.com/2075-4434/8/4/73}.

\bibitem[{\citenamefont{{Sakharov}}(1984)}]{1984JETP...60..214S}
\bibinfo{author}{\bibfnamefont{A.~D.} \bibnamefont{{Sakharov}}},
  \bibinfo{journal}{Soviet Journal of Experimental and Theoretical Physics}
  \textbf{\bibinfo{volume}{60}}, \bibinfo{pages}{214} (\bibinfo{year}{1984}).

\bibitem[{\citenamefont{{Das} et~al.}(1997)\citenamefont{{Das}, {Tariq},
  {Aruliah}, and {Biech}}}]{1997JMP....38.4202D}
\bibinfo{author}{\bibfnamefont{A.}~\bibnamefont{{Das}}},
  \bibinfo{author}{\bibfnamefont{N.}~\bibnamefont{{Tariq}}},
  \bibinfo{author}{\bibfnamefont{D.}~\bibnamefont{{Aruliah}}},
  \bibnamefont{and} \bibinfo{author}{\bibfnamefont{T.}~\bibnamefont{{Biech}}},
  \bibinfo{journal}{Journal of Mathematical Physics}
  \textbf{\bibinfo{volume}{38}}, \bibinfo{pages}{4202} (\bibinfo{year}{1997}).

\bibitem[{\citenamefont{{Hirayama} and {Holdom}}(2003)}]{2003PhRvD..68d4003H}
\bibinfo{author}{\bibfnamefont{T.}~\bibnamefont{{Hirayama}}} \bibnamefont{and}
  \bibinfo{author}{\bibfnamefont{B.}~\bibnamefont{{Holdom}}},
  \bibinfo{journal}{Phys. Rev. D} \textbf{\bibinfo{volume}{68}},
  \bibinfo{eid}{044003} (\bibinfo{year}{2003}), \eprint{hep-th/0303174}.

\bibitem[{\citenamefont{Zhang}(2019{\natexlab{a}})}]{Zhang:2019rrc}
\bibinfo{author}{\bibfnamefont{F.}~\bibnamefont{Zhang}},
  \bibinfo{journal}{Physical Review D} \textbf{\bibinfo{volume}{100}},
  \bibinfo{pages}{064043} (\bibinfo{year}{2019}{\natexlab{a}}),
  \eprint{1909.10669}.

\bibitem[{\citenamefont{{Capozziello} et~al.}(2024)\citenamefont{{Capozziello},
  {De Bianchi}, and {Battista}}}]{2024PhRvD.109j4060C}
\bibinfo{author}{\bibfnamefont{S.}~\bibnamefont{{Capozziello}}},
  \bibinfo{author}{\bibfnamefont{S.}~\bibnamefont{{De Bianchi}}},
  \bibnamefont{and}
  \bibinfo{author}{\bibfnamefont{E.}~\bibnamefont{{Battista}}},
  \bibinfo{journal}{Physical Review D} \textbf{\bibinfo{volume}{109}},
  \bibinfo{eid}{104060} (\bibinfo{year}{2024}), \eprint{2404.17267}.

\bibitem[{\citenamefont{{Hartle} and {Hawking}}(1983)}]{1983PhRvD..28.2960H}
\bibinfo{author}{\bibfnamefont{J.~B.} \bibnamefont{{Hartle}}} \bibnamefont{and}
  \bibinfo{author}{\bibfnamefont{S.~W.} \bibnamefont{{Hawking}}},
  \bibinfo{journal}{Phys. Rev. D} \textbf{\bibinfo{volume}{28}},
  \bibinfo{pages}{2960} (\bibinfo{year}{1983}).

\bibitem[{\citenamefont{{Tod}}(2010)}]{2010JPhCS.229a2013T}
\bibinfo{author}{\bibfnamefont{P.}~\bibnamefont{{Tod}}}, in
  \emph{\bibinfo{booktitle}{Journal of Physics Conference Series}}
  (\bibinfo{year}{2010}), vol. \bibinfo{volume}{229} of
  \emph{\bibinfo{series}{Journal of Physics Conference Series}}, p.
  \bibinfo{pages}{012013}.

\bibitem[{\citenamefont{Zhang}(2019{\natexlab{b}})}]{galaxies7010027}
\bibinfo{author}{\bibfnamefont{F.}~\bibnamefont{Zhang}},
  \bibinfo{journal}{Galaxies} \textbf{\bibinfo{volume}{7}}, \bibinfo{pages}{27}
  (\bibinfo{year}{2019}{\natexlab{b}}), ISSN \bibinfo{issn}{2075-4434},
  \urlprefix\url{http://www.mdpi.com/2075-4434/7/1/27}.

\bibitem[{\citenamefont{{DESI Collaboration} et~al.}(2024)\citenamefont{{DESI
  Collaboration}, {Adame}, {Aguilar}, {Ahlen}, {Alam}, {Alexander}, {Alvarez},
  {Alves}, {Anand}, {Andrade} et~al.}}]{2024arXiv240403002D}
\bibinfo{author}{\bibnamefont{{DESI Collaboration}}},
  \bibinfo{author}{\bibfnamefont{A.~G.} \bibnamefont{{Adame}}},
  \bibinfo{author}{\bibfnamefont{J.}~\bibnamefont{{Aguilar}}},
  \bibinfo{author}{\bibfnamefont{S.}~\bibnamefont{{Ahlen}}},
  \bibinfo{author}{\bibfnamefont{S.}~\bibnamefont{{Alam}}},
  \bibinfo{author}{\bibfnamefont{D.~M.} \bibnamefont{{Alexander}}},
  \bibinfo{author}{\bibfnamefont{M.}~\bibnamefont{{Alvarez}}},
  \bibinfo{author}{\bibfnamefont{O.}~\bibnamefont{{Alves}}},
  \bibinfo{author}{\bibfnamefont{A.}~\bibnamefont{{Anand}}},
  \bibinfo{author}{\bibfnamefont{U.}~\bibnamefont{{Andrade}}},
  \bibnamefont{et~al.}, \bibinfo{journal}{arXiv e-prints}
  \bibinfo{eid}{arXiv:2404.03002} (\bibinfo{year}{2024}), \eprint{2404.03002}.

\bibitem[{\citenamefont{{Poisson} and {Israel}}(1988)}]{1988CQGra...5L.201P}
\bibinfo{author}{\bibfnamefont{E.}~\bibnamefont{{Poisson}}} \bibnamefont{and}
  \bibinfo{author}{\bibfnamefont{W.}~\bibnamefont{{Israel}}},
  \bibinfo{journal}{Classical and Quantum Gravity}
  \textbf{\bibinfo{volume}{5}}, \bibinfo{pages}{L201} (\bibinfo{year}{1988}).

\bibitem[{\citenamefont{{Dymnikova}}(1992)}]{1992GReGr..24..235D}
\bibinfo{author}{\bibfnamefont{I.}~\bibnamefont{{Dymnikova}}},
  \bibinfo{journal}{General Relativity and Gravitation}
  \textbf{\bibinfo{volume}{24}}, \bibinfo{pages}{235} (\bibinfo{year}{1992}).

\bibitem[{\citenamefont{{Markov}}(1982)}]{Markov1982}
\bibinfo{author}{\bibfnamefont{M.~A.} \bibnamefont{{Markov}}},
  \bibinfo{journal}{Sov. Phys. JETP Lett.} \textbf{\bibinfo{volume}{36}},
  \bibinfo{pages}{265} (\bibinfo{year}{1982}).

\bibitem[{\citenamefont{{Dymnikova}}(2006)}]{2006PhLB..639..368D}
\bibinfo{author}{\bibfnamefont{I.}~\bibnamefont{{Dymnikova}}},
  \bibinfo{journal}{Physics Letters B} \textbf{\bibinfo{volume}{639}},
  \bibinfo{pages}{368} (\bibinfo{year}{2006}), \eprint{hep-th/0607174}.

\bibitem[{\citenamefont{{Hall} and {Rendall}}(1987)}]{1987GReGr..19..771H}
\bibinfo{author}{\bibfnamefont{G.~S.} \bibnamefont{{Hall}}} \bibnamefont{and}
  \bibinfo{author}{\bibfnamefont{A.~D.} \bibnamefont{{Rendall}}},
  \bibinfo{journal}{General Relativity and Gravitation}
  \textbf{\bibinfo{volume}{19}}, \bibinfo{pages}{771} (\bibinfo{year}{1987}).

\bibitem[{\citenamefont{{Tkachev} et~al.}(1998)\citenamefont{{Tkachev},
  {Khlebnikov}, {Kofman}, and {Linde}}}]{1998PhLB..440..262T}
\bibinfo{author}{\bibfnamefont{I.}~\bibnamefont{{Tkachev}}},
  \bibinfo{author}{\bibfnamefont{S.}~\bibnamefont{{Khlebnikov}}},
  \bibinfo{author}{\bibfnamefont{L.}~\bibnamefont{{Kofman}}}, \bibnamefont{and}
  \bibinfo{author}{\bibfnamefont{A.}~\bibnamefont{{Linde}}},
  \bibinfo{journal}{Physics Letters B} \textbf{\bibinfo{volume}{440}},
  \bibinfo{pages}{262} (\bibinfo{year}{1998}), \eprint{hep-ph/9805209}.

\bibitem[{\citenamefont{{Sakharov}}(1966)}]{1966JETP...22..241S}
\bibinfo{author}{\bibfnamefont{A.~D.} \bibnamefont{{Sakharov}}},
  \bibinfo{journal}{Soviet Journal of Experimental and Theoretical Physics}
  \textbf{\bibinfo{volume}{22}}, \bibinfo{pages}{241} (\bibinfo{year}{1966}).

\bibitem[{\citenamefont{{Gliner}}(1966)}]{1966JETP...22..378G}
\bibinfo{author}{\bibfnamefont{E.~B.} \bibnamefont{{Gliner}}},
  \bibinfo{journal}{Soviet Journal of Experimental and Theoretical Physics}
  \textbf{\bibinfo{volume}{22}}, \bibinfo{pages}{378} (\bibinfo{year}{1966}).

\bibitem[{\citenamefont{{Frolov} et~al.}(1990)\citenamefont{{Frolov}, {Markov},
  and {Mukhanov}}}]{1990PhRvD..41..383F}
\bibinfo{author}{\bibfnamefont{V.~P.} \bibnamefont{{Frolov}}},
  \bibinfo{author}{\bibfnamefont{M.~A.} \bibnamefont{{Markov}}},
  \bibnamefont{and} \bibinfo{author}{\bibfnamefont{V.~F.}
  \bibnamefont{{Mukhanov}}}, \bibinfo{journal}{Phys. Rev. D}
  \textbf{\bibinfo{volume}{41}}, \bibinfo{pages}{383} (\bibinfo{year}{1990}).

\bibitem[{\citenamefont{{Balbinot} and {Poisson}}(1990)}]{1990PhRvD..41..395B}
\bibinfo{author}{\bibfnamefont{R.}~\bibnamefont{{Balbinot}}} \bibnamefont{and}
  \bibinfo{author}{\bibfnamefont{E.}~\bibnamefont{{Poisson}}},
  \bibinfo{journal}{Phys. Rev. D} \textbf{\bibinfo{volume}{41}},
  \bibinfo{pages}{395} (\bibinfo{year}{1990}).

\bibitem[{\citenamefont{{Gron}}(1985)}]{1985NCimL..44..177G}
\bibinfo{author}{\bibfnamefont{O.}~\bibnamefont{{Gron}}},
  \bibinfo{journal}{Nuovo Cimento Lettere} \textbf{\bibinfo{volume}{44}},
  \bibinfo{pages}{177} (\bibinfo{year}{1985}).

\bibitem[{\citenamefont{{Madsen}}(1985)}]{1985Ap&SS.113..205M}
\bibinfo{author}{\bibfnamefont{M.~S.} \bibnamefont{{Madsen}}},
  \bibinfo{journal}{Astrophys. Sp. Sci.} \textbf{\bibinfo{volume}{113}},
  \bibinfo{pages}{205} (\bibinfo{year}{1985}).

\bibitem[{\citenamefont{{Bardeen}}(1968)}]{1968Bardeen}
\bibinfo{author}{\bibfnamefont{J.~M.} \bibnamefont{{Bardeen}}},
  \bibinfo{journal}{Proceedings of the International Conference GR5, Tbilisi,
  USSR}  (\bibinfo{year}{1968}).

\bibitem[{\citenamefont{{Mazur} and {Mottola}}(2001)}]{2001gr.qc.....9035M}
\bibinfo{author}{\bibfnamefont{P.~O.} \bibnamefont{{Mazur}}} \bibnamefont{and}
  \bibinfo{author}{\bibfnamefont{E.}~\bibnamefont{{Mottola}}},
  \bibinfo{journal}{arXiv e-prints} \bibinfo{eid}{gr-qc/0109035}
  (\bibinfo{year}{2001}), \eprint{gr-qc/0109035}.

\bibitem[{\citenamefont{{Dymnikova} and {Kraav}}(2019)}]{2019Univ....5..163D}
\bibinfo{author}{\bibfnamefont{I.}~\bibnamefont{{Dymnikova}}} \bibnamefont{and}
  \bibinfo{author}{\bibfnamefont{K.}~\bibnamefont{{Kraav}}},
  \bibinfo{journal}{Universe} \textbf{\bibinfo{volume}{5}},
  \bibinfo{pages}{163} (\bibinfo{year}{2019}).

\bibitem[{\citenamefont{{Pani} et~al.}(2009)\citenamefont{{Pani}, {Berti},
  {Cardoso}, {Chen}, and {Norte}}}]{2009PhRvD..80l4047P}
\bibinfo{author}{\bibfnamefont{P.}~\bibnamefont{{Pani}}},
  \bibinfo{author}{\bibfnamefont{E.}~\bibnamefont{{Berti}}},
  \bibinfo{author}{\bibfnamefont{V.}~\bibnamefont{{Cardoso}}},
  \bibinfo{author}{\bibfnamefont{Y.}~\bibnamefont{{Chen}}}, \bibnamefont{and}
  \bibinfo{author}{\bibfnamefont{R.}~\bibnamefont{{Norte}}},
  \bibinfo{journal}{Phys. Rev. D} \textbf{\bibinfo{volume}{80}},
  \bibinfo{eid}{124047} (\bibinfo{year}{2009}), \eprint{0909.0287}.

\bibitem[{\citenamefont{{Cruz} et~al.}(2017)\citenamefont{{Cruz}, {Ganguly},
  {Gannouji}, {Leon}, and {Saridakis}}}]{2017CQGra..34l5014C}
\bibinfo{author}{\bibfnamefont{M.}~\bibnamefont{{Cruz}}},
  \bibinfo{author}{\bibfnamefont{A.}~\bibnamefont{{Ganguly}}},
  \bibinfo{author}{\bibfnamefont{R.}~\bibnamefont{{Gannouji}}},
  \bibinfo{author}{\bibfnamefont{G.}~\bibnamefont{{Leon}}}, \bibnamefont{and}
  \bibinfo{author}{\bibfnamefont{E.~N.} \bibnamefont{{Saridakis}}},
  \bibinfo{journal}{Classical and Quantum Gravity}
  \textbf{\bibinfo{volume}{34}}, \bibinfo{eid}{125014} (\bibinfo{year}{2017}),
  \eprint{1702.01754}.

\bibitem[{\citenamefont{{Clarkson} and {Barrett}}(2003)}]{2003CQGra..20.3855C}
\bibinfo{author}{\bibfnamefont{C.~A.} \bibnamefont{{Clarkson}}}
  \bibnamefont{and} \bibinfo{author}{\bibfnamefont{R.~K.}
  \bibnamefont{{Barrett}}}, \bibinfo{journal}{Classical and Quantum Gravity}
  \textbf{\bibinfo{volume}{20}}, \bibinfo{pages}{3855} (\bibinfo{year}{2003}),
  \eprint{gr-qc/0209051}.

\bibitem[{\citenamefont{{Fisher}}(1948)}]{Fisher1948}
\bibinfo{author}{\bibfnamefont{I.~Z.} \bibnamefont{{Fisher}}},
  \bibinfo{journal}{Zh. Eksp. Teor. Fiz.} \textbf{\bibinfo{volume}{18}},
  \bibinfo{pages}{636} (\bibinfo{year}{1948}).

\bibitem[{\citenamefont{{Bergmann} and {Leipnik}}(1957)}]{1957PhRv..107.1157B}
\bibinfo{author}{\bibfnamefont{O.}~\bibnamefont{{Bergmann}}} \bibnamefont{and}
  \bibinfo{author}{\bibfnamefont{R.}~\bibnamefont{{Leipnik}}},
  \bibinfo{journal}{Physical Review} \textbf{\bibinfo{volume}{107}},
  \bibinfo{pages}{1157} (\bibinfo{year}{1957}).

\bibitem[{\citenamefont{{Janis} et~al.}(1968)\citenamefont{{Janis}, {Newman},
  and {Winicour}}}]{1968PhRvL..20..878J}
\bibinfo{author}{\bibfnamefont{A.~I.} \bibnamefont{{Janis}}},
  \bibinfo{author}{\bibfnamefont{E.~T.} \bibnamefont{{Newman}}},
  \bibnamefont{and}
  \bibinfo{author}{\bibfnamefont{J.}~\bibnamefont{{Winicour}}},
  \bibinfo{journal}{Phys. Rev. Lett.} \textbf{\bibinfo{volume}{20}},
  \bibinfo{pages}{878} (\bibinfo{year}{1968}).

\bibitem[{\citenamefont{{Wyman}}(1981)}]{1981PhRvD..24..839W}
\bibinfo{author}{\bibfnamefont{M.}~\bibnamefont{{Wyman}}},
  \bibinfo{journal}{Phys. Rev. D} \textbf{\bibinfo{volume}{24}},
  \bibinfo{pages}{839} (\bibinfo{year}{1981}).

\bibitem[{\citenamefont{{Bekenstein}}(1972)}]{1972PhRvL..28..452B}
\bibinfo{author}{\bibfnamefont{J.~D.} \bibnamefont{{Bekenstein}}},
  \bibinfo{journal}{Phys. Rev. Lett.} \textbf{\bibinfo{volume}{28}},
  \bibinfo{pages}{452} (\bibinfo{year}{1972}).

\bibitem[{\citenamefont{{Jebsen}}(2005)}]{2005GReGr..37.2253J}
\bibinfo{author}{\bibfnamefont{J.~T.} \bibnamefont{{Jebsen}}},
  \bibinfo{journal}{General Relativity and Gravitation}
  \textbf{\bibinfo{volume}{37}}, \bibinfo{pages}{2253} (\bibinfo{year}{2005}).

\bibitem[{\citenamefont{{Ganguly} et~al.}(2015)\citenamefont{{Ganguly},
  {Gannouji}, {Goswami}, and {Ray}}}]{2015CQGra..32j5006G}
\bibinfo{author}{\bibfnamefont{A.}~\bibnamefont{{Ganguly}}},
  \bibinfo{author}{\bibfnamefont{R.}~\bibnamefont{{Gannouji}}},
  \bibinfo{author}{\bibfnamefont{R.}~\bibnamefont{{Goswami}}},
  \bibnamefont{and} \bibinfo{author}{\bibfnamefont{S.}~\bibnamefont{{Ray}}},
  \bibinfo{journal}{Classical and Quantum Gravity}
  \textbf{\bibinfo{volume}{32}}, \bibinfo{eid}{105006} (\bibinfo{year}{2015}),
  \eprint{1411.2601}.

\bibitem[{\citenamefont{Strogatz}(2007)}]{strogatz2007nonlinear}
\bibinfo{author}{\bibfnamefont{S.}~\bibnamefont{Strogatz}},
  \emph{\bibinfo{title}{Nonlinear Dynamics and Chaos: With Applications to
  Physics, Biology, Chemistry, and Engineering}}, Studies in nonlinearity
  (\bibinfo{publisher}{Levant Books}, \bibinfo{year}{2007}), ISBN
  \bibinfo{isbn}{9788187169857},
  \urlprefix\url{https://books.google.co.jp/books?id=PHmED2xxrE8C}.

\bibitem[{\citenamefont{{Poisson} and {Israel}}(1989)}]{1989PhRvL..63.1663P}
\bibinfo{author}{\bibfnamefont{E.}~\bibnamefont{{Poisson}}} \bibnamefont{and}
  \bibinfo{author}{\bibfnamefont{W.}~\bibnamefont{{Israel}}},
  \bibinfo{journal}{Phys. Rev. Lett.} \textbf{\bibinfo{volume}{63}},
  \bibinfo{pages}{1663} (\bibinfo{year}{1989}).

\bibitem[{\citenamefont{{Poisson} and {Israel}}(1990)}]{1990PhRvD..41.1796P}
\bibinfo{author}{\bibfnamefont{E.}~\bibnamefont{{Poisson}}} \bibnamefont{and}
  \bibinfo{author}{\bibfnamefont{W.}~\bibnamefont{{Israel}}},
  \bibinfo{journal}{Phys. Rev. D} \textbf{\bibinfo{volume}{41}},
  \bibinfo{pages}{1796} (\bibinfo{year}{1990}).

\bibitem[{\citenamefont{{Chambers}}(1997)}]{1997AnIPS..13...33C}
\bibinfo{author}{\bibfnamefont{C.~M.} \bibnamefont{{Chambers}}},
  \bibinfo{journal}{Annals of the Israel Physical Society}
  \textbf{\bibinfo{volume}{13}}, \bibinfo{pages}{33} (\bibinfo{year}{1997}),
  \eprint{gr-qc/9709025}.

\bibitem[{\citenamefont{{Brady} et~al.}(1998)\citenamefont{{Brady}, {Moss}, and
  {Myers}}}]{1998PhRvL..80.3432B}
\bibinfo{author}{\bibfnamefont{P.~R.} \bibnamefont{{Brady}}},
  \bibinfo{author}{\bibfnamefont{I.~G.} \bibnamefont{{Moss}}},
  \bibnamefont{and} \bibinfo{author}{\bibfnamefont{R.~C.}
  \bibnamefont{{Myers}}}, \bibinfo{journal}{\prl}
  \textbf{\bibinfo{volume}{80}}, \bibinfo{pages}{3432} (\bibinfo{year}{1998}),
  \eprint{gr-qc/9801032}.

\bibitem[{\citenamefont{Ori}(1991)}]{PhysRevLett.67.789}
\bibinfo{author}{\bibfnamefont{A.}~\bibnamefont{Ori}}, \bibinfo{journal}{Phys.
  Rev. Lett.} \textbf{\bibinfo{volume}{67}}, \bibinfo{pages}{789}
  (\bibinfo{year}{1991}),
  \urlprefix\url{https://link.aps.org/doi/10.1103/PhysRevLett.67.789}.

\bibitem[{\citenamefont{{Burko} and {Ori}}(1998)}]{1998PhRvD..57.7084B}
\bibinfo{author}{\bibfnamefont{L.~M.} \bibnamefont{{Burko}}} \bibnamefont{and}
  \bibinfo{author}{\bibfnamefont{A.}~\bibnamefont{{Ori}}},
  \bibinfo{journal}{Phys. Rev. D} \textbf{\bibinfo{volume}{57}},
  \bibinfo{pages}{R7084} (\bibinfo{year}{1998}), \eprint{gr-qc/9711032}.

\end{thebibliography}

\end{document}